\documentclass[aps,prb,reprint,amsmath,superscriptaddress,amssymb,showpacs]{revtex4-1}
\usepackage[english]{babel}
\usepackage{graphicx}
\usepackage[utf8]{inputenc}
\usepackage{color}

\makeatletter
\renewcommand{\fnum@figure}{FIG. \thefigure}
\makeatother

\newcommand{\D}{\mathcal{D}}
\renewcommand{\P}{\mathcal{P}}
\newcommand{\um}[1]{\,\rm{#1}}

\newcommand{\beq}{\begin{equation}}
\newcommand{\eeq}{\end{equation}}
\newcommand{\be}{\begin{equation}}
\newcommand{\ee}{\end{equation}}
\newcommand{\beqa}{\begin{eqnarray}}
\newcommand{\eeqa}{\end{eqnarray}}
\newcommand{\bea}{\begin{align}}
\newcommand{\eea}{\end{align}}

\newcommand{\vk}{\mathbf{k}}
\newcommand{\vq}{\mathbf{q}}
\newcommand{\St}[1]{\textrm{St}_{\textrm{#1}}}
\def\Xint#1{\mathchoice
   {\XXint\displaystyle\textstyle{#1}}%
   {\XXint\textstyle\scriptstyle{#1}}%
   {\XXint\scriptstyle\scriptscriptstyle{#1}}%
   {\XXint\scriptscriptstyle\scriptscriptstyle{#1}}%
   \!\int}
\def\XXint#1#2#3{{\setbox0=\hbox{$#1{#2#3}{\int}$}
     \vcenter{\hbox{$#2#3$}}\kern-.5\wd0}}

\def\dashint{\Xint-}

\begin{document}

\title{Negative absolute conductivity in photoexcited metals}
\author{Giuliano Chiriacò}
\affiliation{Department of Physics, Columbia University, New York, New York 10027, USA}
\author{Andrew J. Millis}
\affiliation{Department of Physics, Columbia University, New York, New York 10027, USA}
\affiliation{Center for Computational Quantum Physics, The Flatiron Institute, New York, New York 10010, USA}
\author{Igor L. Aleiner}
\affiliation{Department of Physics, Columbia University, New York, New York 10027, USA}

\date{\today}

\begin{abstract}
We show that in a model of a metal photoexcited by a transient pump pulse resonant with a phonon mode, the absolute dc conductivity may become negative, depending on the interplay between the electronic structure, the phonon frequency and the pump intensity. The analysis includes the effects of inelastic scattering and thermal relaxation. Results for the time evolution of the negative conductivity state are presented; the associated non-equilibrium physics may persist for long times after the pulse. Our findings provide a theoretical justification for previously proposed phenomenology and indicate new routes to the generation and exploration of intrinsically non-equilibrium states.
\end{abstract}

\maketitle

The dc electrical conductivity $\sigma$ (ratio of current $j$ to applied field $E$) is a fundamental property of materials. In thermal equilibrium the linear response conductivity is non-negative because an applied electric field creates entropy via  Joule heating $\sigma E^2$ and the entropy production rate must be non-negative. Beyond the linear response regime new effects may occur. For example, a negative {\em differential }conductivity $\sigma_{\textrm{diff}}\equiv \textrm{d}j/\textrm{d}E_{|E\neq0}<0$ has been extensively studied \cite{Zakh,*Zakh2,Volk,Ridl,*Ridl2} and is typically related to runaway heating at current driven metal-insulator transitions. This paper is concerned with the less commonly realized situation of negative {\em absolute} conductivity (NAC), $\sigma\equiv j/E<0$. A negative absolute conductivity state is possible away from thermal equilibrium  because the entropy decrease implied by the $\sigma E^2$ term can be compensated by other sources of entropy production, and would lead to remarkable phenomenological consequences including novel response properties \cite{zerodc:exp1,zerodc:exp2,Rizhi,*Rizhi2} spontaneously generated internal electric fields \cite{IA:AJM},  and new collective modes \cite{CMA} that might be relevant to recent experimental studies of the transient optical properties in photoexcited K$_3$C$_{60}$ \cite{K3:C60}. It is therefore important to understand the circumstances under which a negative absolute conductivity can occur.

Insight into the origin of the  NAC state may be obtained from the expression $\sigma=\int d\varepsilon\tilde{\sigma}(\varepsilon)(-\partial_{\varepsilon}f)$, with $\tilde{\sigma}(\varepsilon)=e^2\langle v^2(\varepsilon)\rangle D(\varepsilon)\tau_{\textrm{tr}}(\varepsilon)$, where $e$ is the electron charge, $\langle v^2\rangle$ is a suitably averaged electron velocity, $D$ is the density of states, $\tau_{\textrm{tr}}$ is the transport scattering time, and $f$ is the electron distribution function. $\tilde{\sigma}$ is always positive and in equilibrium $-\partial_{\varepsilon}f>0$. However, out of equilibrium $-\partial_{\varepsilon}f$ may become negative in some energy regions; we refer to this situation as a local (in energy) population inversion. If the energy regions where $-\partial_{\varepsilon}f<0$ coincide with maxima of $\tilde{\sigma}$, then the total conductivity may become negative. Regions of $-\partial_{\varepsilon}f<0$ were shown to occur and to lead to negative absolute conductivity in the two-dimensional electron gas subject to a perpendicular magnetic field and to a steady state microwave radiation \cite{Sach,Dyak,*Dyak2,Vavilov}, and more recently in a steadily photoexcited correlated insulator \cite{Aoki} and in a thermally driven SIS junction \cite{Marche}; for example, in the first system the peaked energy structure in $\tilde{\sigma}$ was caused by Landau level quantization and the regions of local inversion were produced by the drive at a frequency that matched the Landau level spacing.

In this paper we show that a local population inversion can occur in a system of electrons coupled to strongly pumped phonons, and that this inversion can lead to a NAC state, even when the pumping is not continuous; indeed, the effect can be induced by transiently pumped phonons and can persist for long times after the pump is removed. We show how the NAC state depends on the intensity of the driving pump and that the effect is maximized if the phonon frequency is approximately commensurate with the distance from Fermi energy to the band edges; we provide information on which forms of the electron and phonon density of states create the most likely conditions for the effect to occur. We estimate the coupling constant from the phenomenological theory of Ref.~\cite{CMA} and  explicate the effects of internal electric fields and energy relaxation mechanisms.

\textit{The model} - We study a metallic system, initially in equilibrium at temperature $T$, characterized by a dispersionless phonon mode with energy $\omega_p$; a weak dispersion is important, as discussed below. We assume (as in the usual theory of electron-phonon coupling) that the electrons and phonons can be described in a quasiparticle picture. Introducing the operators $c_{\vk},c^{\dag}_{\vk}$ and $a_{\vq},a_{\vq}^{\dag}$ for electron and phonons respectively, the Hamiltonian can be written as
\begin{gather}\label{Ham}
H=\sum_{\vk}\epsilon_{\vk}c^{\dagger}_{\vk}c_{\vk}+
\sum_{\vq}\omega_pa^{\dagger}_{\vq}a_{\vq}+H_{\textrm{el-ph}};
\end{gather}
where $H_{\textrm{el-ph}}=\sum_{\vk,\vq}M_{\vq}(a^{\dagger}_{-\vq}+a_{\vq})c^{\dagger}_{\vk}c_{\vk-\vq}$, with $M_{\vq}$ the electron-phonon interaction matrix element, and $\epsilon_{\vk}$ is the electron energy dispersion.

We assume that the system is photoexcited by radiation that induces a highly non-equilibrium state of the phonons and we assume that the phonon coherence and momentum relax very quickly, so we may characterize the non-equilibrium phonon population by a diagonal, momentum-independent distribution function $\langle a^{\dag}_{\vq}a_{\vq}\rangle=\zeta+b$, which is the sum of the thermal distribution $b=(e^{\omega_p/T}-1)^{-1}$ and a non equilibrium component $\zeta$. Because of the momentum independence of $\zeta$ we can average all the relevant electronic properties over $\vk$ and characterize the system by $\zeta$, the electron distribution $f(\varepsilon)$, the density of states $D(\varepsilon)$, the average velocity squared $v^2(\varepsilon)$ and the transport scattering time $\tau_{\textrm{tr}}$ \footnote{For the purpose of numerical calculations, we use either $D(\varepsilon)$ and $v^2(\varepsilon)$ derived from a model of p-like bands or a trial $D(\varepsilon)$, but the origin is not relevant, since only the $\varepsilon$ structure is important.}.

We study the non-equilibrium dynamics of the system using the Keldysh formalism within Migdal-Eliashberg theory; the supplemental material provides a detailed treatment. We find as in equilibrium that the electron density of states (retarded part of the Green function) and the phonon frequency are only slightly renormalized by the non-equilibrium drive \cite{supp}. We therefore focus on the electron distribution function $f$ and on the non-equilibrium part of the phonon population $\zeta$, which are the solution of two coupled kinetic equations:
\begin{gather}\label{KEf}
\partial_tf+\textrm{St}_{\textrm{E}}\{f\}=\textrm{St}_{\textrm{in}}\{f\}+\textrm{St}_{\textrm{el}}\{f,\zeta\};\\
\label{Kez}\partial_t\zeta=\textrm{St}_{\textrm{ph}}\{f,\zeta\}+I_p(t)-\zeta/\tau_{\textrm{ph}},
\end{gather}
where $\textrm{St}_{\textrm{E}}$ is the effect of the dc electric field $E$, $\textrm{St}_{\textrm{in}}$ is the inelastic scattering term, $\textrm{St}_{\textrm{el}}$ and $\textrm{St}_{\textrm{ph}}$ are the contributions of the electron-phonon interaction to the collision integrals of $f$ and of $\zeta$ respectively, $I_p(t)$ a phonon source term arising from the pump and the initial decoherence processes, and $\tau_{\textrm{ph}}$ is the decay time for $\zeta$, due to inelastic scattering with other phonons \cite{supp}. Notice that in general the pump pulse also affects the electrons, but it has essentially the same effects a phonons, since it drives the same electronic transitions; for simplicity we neglect this effect, since it would not affect the steady state electronoc distribution and would just accelerate the initial evolution of the electrons in the transient regime.

Neglecting for simplicity the $\vq$ dependence of $M_{\vq}$, we evaluate the collision integral for electrons and phonons
\begin{align}\label{Stel}
\notag \textrm{St}_{\textrm{el}}=&\frac{\Gamma_{eph}}{D_0}\Big(
D_{\varepsilon-\omega_p}[(\zeta+b)(f_{\varepsilon-\omega_p}-f_{\varepsilon})-f_{\varepsilon}(1-f_{\varepsilon-\omega_p})]+\\
+&
D_{\varepsilon+\omega_p}[(\zeta+b)(f_{\varepsilon+\omega_p}-f_{\varepsilon})+f_{\varepsilon+\omega_p}(1-f_{\varepsilon})]\Big)\\
\label{Stph}
\notag \textrm{St}_{\textrm{ph}}&=\frac{\Gamma_{eph}}{D_0}\int D(\varepsilon)D(\varepsilon+\omega_p)\Big[f(\varepsilon+\omega_p)(1-f(\varepsilon))+\\
&+(\zeta+b)\Big(f(\varepsilon+\omega_p)-f(\varepsilon)\Big)\Big] d\varepsilon,
\end{align}
where $\Gamma_{eph}\equiv2\pi|M|^2D_0$ is the electron-phonon scattering rate and $D_0$ is the average electron density of states. Equation \eqref{Stel} has an evident periodicity in energy, which at $\zeta\gg1$ induces a periodic distribution $f(\varepsilon)$ with period $\omega_p$; for such distribution, both $\St{el}$ and $\St{ph}$ approximately vanish.

We model the inelastic scattering as arising from the coupling to a thermal bath at temperature $T$; if the energy is exchanged in small amounts, the scattering is an energy diffusion process with effective rate $\Gamma_{in}$:
\begin{align}\label{Stin}
\textrm{St}_{\textrm{in}}=\frac{\Gamma_{in}}{D_F}\frac1{D(\varepsilon)}\partial_{\varepsilon}\Big[
D^2(\varepsilon)[T\partial_{\varepsilon}f+f(1-f)]\Big],
\end{align}
$\St{in}$ makes the electrons relax to a Fermi-Dirac distribution with temperature $T$.

It will also be important to consider an applied dc electric field. As shown in the supplemental material, this causes a diffusion in energy space
\begin{equation}\label{STE}
\textrm{St}_{\textrm{E}}=-\frac{E^2}3\frac{1}{D(\varepsilon)}\partial_{\varepsilon}
[\tilde{\sigma}(\varepsilon)\partial_{\varepsilon}f(\varepsilon)],
\end{equation}
where $\tilde{\sigma}(\varepsilon)=e^2v^2(\varepsilon)D(\varepsilon)\tau_{\textrm{tr}}$. We see from Eq. \eqref{STE} that the electric field smooths out the steepest regions in $f$, creating a pseudo-thermal distribution \cite{CM:cdw,Han:cdw} with effective temperature $T_{\textrm{eff}}\sim T+e^2E^2v_F^2\tau_{\textrm{tr}}/\Gamma_{in}$, where $v_F$ is the Fermi velocity.

Equations \eqref{KEf}-\eqref{STE} are a complete system that can be solved for $f(\varepsilon,t)$ and $\zeta(t)$ given a source term $I_p(t)$. We consider two limiting cases: \textit{i)} a steady state drive; \textit{ii)} a short pump pulse occurring over a time $\tau_{\textrm{pulse}}$ much smaller than the relaxation time of the transient state.

\textit{Population inversion for steady state drive} - In equilibrium ($\zeta=0$, $E=0$) Eq. \eqref{KEf} is solved by the thermal Fermi-Dirac distribution $f_T(\varepsilon)$. To gain a first understanding of the non-equilibrium physics, we neglect inelastic scattering of electrons ($\St{in}\rightarrow0$), electric field and phonon dynamics; we assume the system to be in equilibrium at temperature $T$ for $t<0$ and that at $t=0$ the phonon distribution is instantaneously switched to a state with $\zeta>0$. We then solve Eq. \eqref{KEf} for fixed $\zeta$ and consider the long time limit.

The dispersionless phonon approximation means that an electronic state at energy $\varepsilon$ is coupled to the discrete set of states at energy $\varepsilon+j\omega_p$, with $j$ an integer such that $\varepsilon+j\omega_p$ is within the band of allowed states. Since the scattering conserves particles number, $\sum_jD(\varepsilon+j\omega_p)f(\varepsilon+j\omega_p)$ is time independent and thus equal to the initial value $\sum_jD(\varepsilon+j\omega_p)f_T(\varepsilon+j\omega_p)$. In the large $\zeta$ limit, $f(\varepsilon)$ must be periodic in $\varepsilon$ so that $\St{el}=0$, i.e. $f(\varepsilon+\omega_p)=f(\varepsilon)$, implying
\begin{equation}\label{fneq}
f(\varepsilon)=\frac{\sum_jD(\varepsilon+j\omega_p)f_T(\varepsilon+j\omega_p)}{\sum_jD(\varepsilon+j\omega_p)}
\end{equation}
\begin{figure}[t]
\centering
\includegraphics[width=0.96\columnwidth]{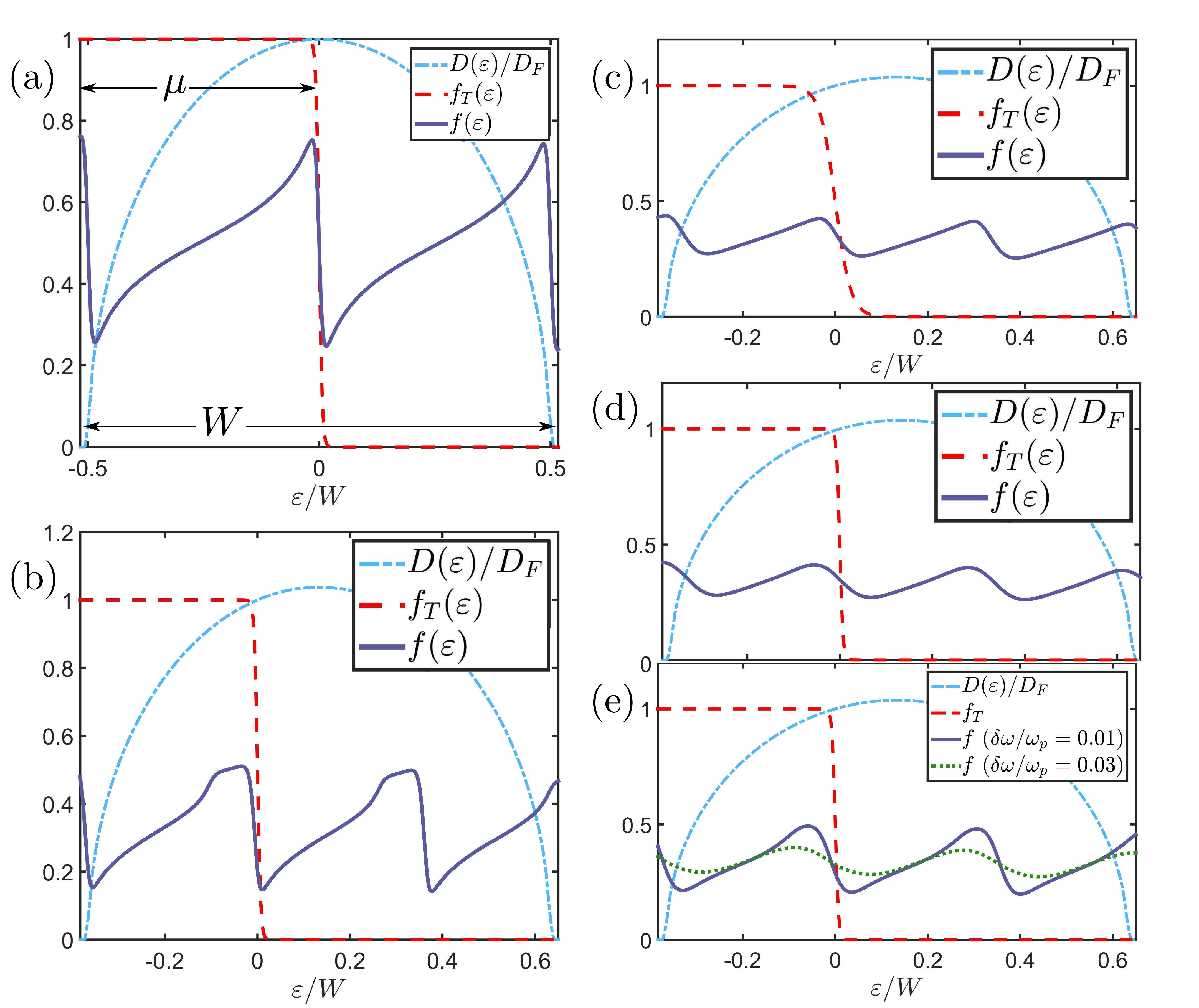}
\caption{\footnotesize{Non-equilibrium steady state electron distribution $f$ (blue solid lines), obtained from solution of Eq. \eqref{KEf} for a steady state phonon population $\zeta=20$, trial DoS $D(\varepsilon)$ (cyan dashed-dotted lines) normalized to the Fermi DoS $D_F$, initial distribution given by a Fermi-Dirac $f_T$ (red dashed) at chemical potential $\mu$ and temperature $T/W=0.003$ (a), (b), (d), (e) and $T/W=0.02$ (c). The phonon frequency is $\omega_p/W=0.5$ (a) and $\omega_p/W=0.36$ (b)-(e). Panel (c) includes a stronger inelastic scattering $\St{in}/\St{el}\sim0.05$; in panel (d) we use the parameters of (b) but with field $eEv_F\sqrt{\tau_{\textrm{tr}}/\Gamma_{in}\omega_p}=0.4$; in panel (e) we use the parameters of (b) and a dispersive phonon with typical width $\delta\omega/\omega_p=0.01$, $0.03$.}}
\label{F1}
\end{figure}

The particular shape of $f(\varepsilon)$ depends on the density of states (DoS) and on $\omega_p$. In the $T\rightarrow0$ limit, the $\varepsilon$ structure of $f$ is controlled by the energy dependence of $D$ in the range between the chemical potential $\mu$ and the lower band edge, except for down steps at $\varepsilon+j\omega_p=\mu$ or steps of either sign when $\varepsilon+j\omega_p$ matches a singularity in the DoS. Since $f$ is periodic, the down steps must be matched by an average increase of $f$.

Results of a numerical solution of Eq. \eqref{KEf} are shown in Fig. 1 for a trial density of states. Here $D$ is an increasing function of $\varepsilon$ between the lower band edge and $\mu$ and we see that $f$ is characterized by regions of smooth increase separated by downward jumps at $\varepsilon=\mu-j\omega_p$; the distribution arising from an alternative DoS (with singularities at the band edges) is shown in the supplement. Panels (a) and (b) of Fig.1 show the $\St{in}\rightarrow0$ limit at different doping levels. Panels (c) and (d) show the effects of including the inelastic scattering (c) and a dc electric field (d); both these terms lead to diffusion in energy space, smoothing out $f$ similarly to raising $T$.

We also analyze the consequences of a dispersive phonon frequency with typical width $\delta\omega$. This leads to an additional diffusion-like term in $\St{el}$ \cite{supp}, which renormalizes the temperature $T\rightarrow T_{\textrm{eff}}=T+\zeta\delta\omega$ and smooths the local population inversion when $T_{\textrm{eff}}\approx\omega_p$, i.e. $\delta\omega/\omega_p\gtrsim1/\zeta$ (Fig. 1e). In the rest of the paper, we neglect the effects of a dispersive band, but allow for a small inelastic scattering and for non zero dc fields.

\textit{Steady state conductivity} - An analysis of the Keldysh equations yields for the conductivity \cite{supp}
\begin{equation}\label{sigma}
\sigma=\int\tilde{\sigma}(\varepsilon)\left(-\partial f/\partial\varepsilon\right) d\varepsilon.
\end{equation}
The sign of $\sigma$ depends on how regions with large and small values of $\tilde{\sigma}(\varepsilon)$ are matched to the regions of normal and inverted population. An expression for $\sigma$ can be derived by approximating $-\partial_{\varepsilon}f$ as the sum of delta functions at $\varepsilon=\mu+j\omega_p$ and smooth terms; for the DoS of Fig. 1, $-\partial_{\varepsilon}f\sim-1/\omega_p$ and we obtain
\begin{equation}\label{sigappr}
\sigma\sim\sum_j\tilde{\sigma}(\mu+j\omega_p)-\frac1{\omega_p}\int d\varepsilon \tilde{\sigma}(\varepsilon)
\end{equation}
From Eq. \eqref{sigappr} we see that when $\omega_p$ is such that $\mu+j\omega_p$ corresponds to a band edge (where $\tilde{\sigma}(\varepsilon)$ is small) for some $j$, the positive term in $\sigma$ may be outweighed by the negative contribution of the integral. This is most likely to happen when $\omega_p$ is commensurate with the distance of either of the band edge energies from the chemical potential, as confirmed by numerical calculations of $\sigma(\omega_p)$ performed in the limit of constant $v^2\tau_{\textrm{tr}}$ \footnote{Alternative choices for transport are possible: for example $\tau_{\textrm{tr}}(\varepsilon)\propto1/D(\varepsilon)$ as for impurity scattering, or $v(\varepsilon)\tau_{\textrm{tr}}(\varepsilon)\sim\text{const}$ as for hard sphere scattering. The results are qualitatively equivalent, with only slight quantitative differences.}, see Fig. 2; indeed the effect is enhanced when the chemical potential is such that $\omega_p$ is commensurate with both band edges energies at the same time, see Fig. 1a and 2a (1/2-filling). A similar criterion holds for more complicated density of states, such as a double peaked structure modeling $p$-like electrons in a cubic lattice; in this case $\sigma(\omega_p)<0$ also when $\omega_p$ is commensurate with the distance from Fermi level to the minimum of $D(\varepsilon)$ (Fig. 2c).
\begin{figure}[t]
\centering
\includegraphics[width=0.96\columnwidth]{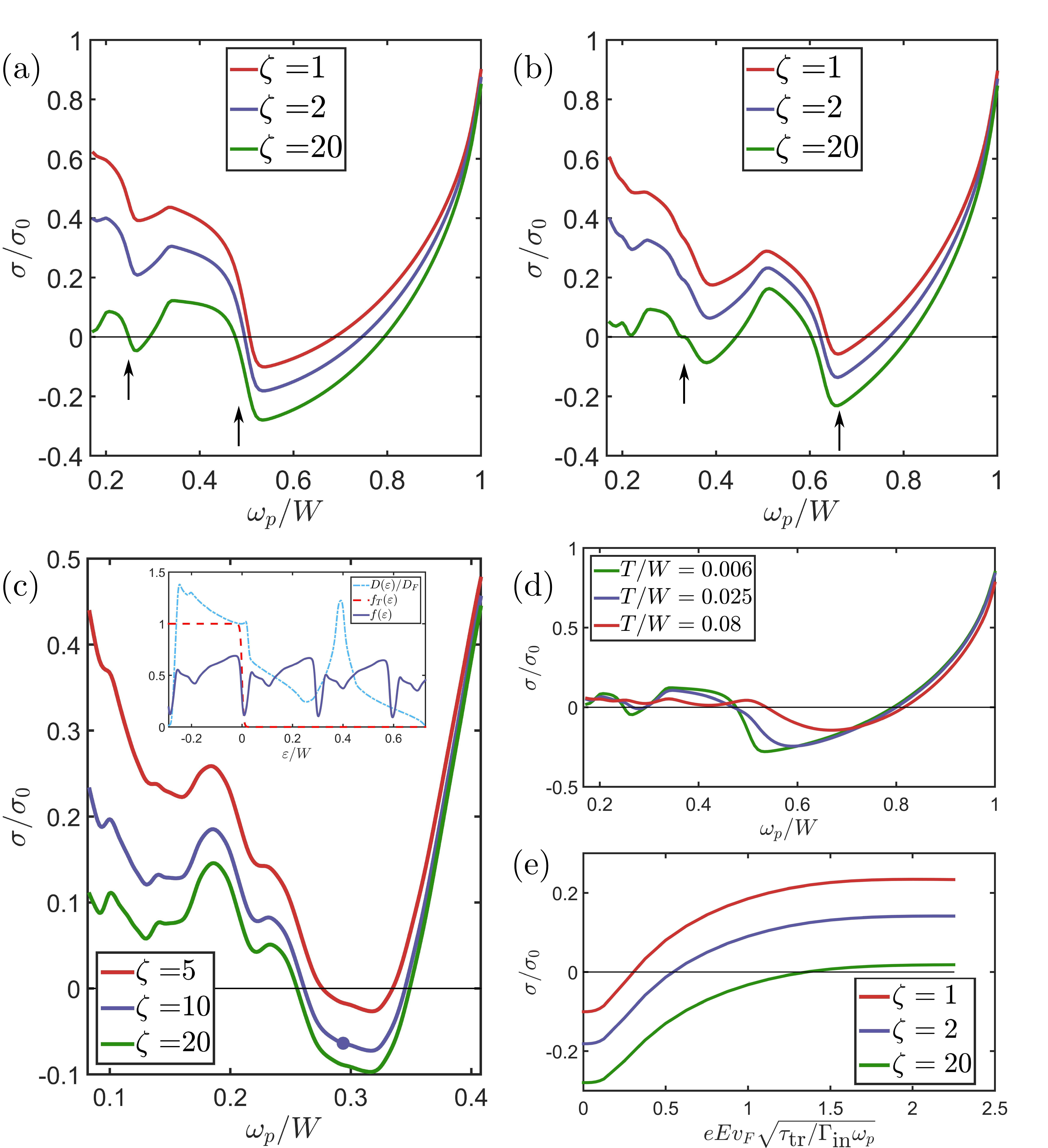}
\caption{\footnotesize{(a)-(b) Plot of normalized conductivity $\sigma/\sigma_0$ (where $\sigma_0\equiv v_F^2\tau_{\textrm{tr}}D_F$) as function of $\omega_p$ for three values of $\zeta$ at $T/W=0.003$, $E=0$ for the DoS of Fig.1; the filling is $1/2$ ($\mu=0$) in (a) and $1/3$ ($\mu\approx -W/6$) in (b); the arrows indicate the values of $\omega_p$ corresponding to the commensurability criteria, i.e. $\omega_p/W=1/4$, $1/2$ in (a) and $\omega_p/W=1/3$, $2/3$ in (b). (c) Plot of $\sigma/\sigma_0$ for a different DoS (modeling $p$-like electrons in cubic symmetry) as function of $\omega_p$ at $T/W=0.003$ and half filling; the inset shows the corresponding DoS $D$ and distribution $f$ for the frequency $\omega_p$ marked with a dot on the graph. (d)-(e) Plot of $\sigma/\sigma_0$ at half filling as function of $\omega_p$ at $\zeta=20$ and $E=0$ for three different temperatures (d) and as function of the normalized electric field $E$ at $\omega_p/W=0.53$ and $T/W=0.003$ (e). Calculations were performed for $\Gamma_{in}\ll\Gamma_{eph}$ assuming constant $v^2\tau_{\textrm{tr}}$ and the system was evolved for a time $10\Gamma_{eph}^{-1}$.}}
\label{F2}
\end{figure}

Figure 2 shows that when plotted as function of the phonon frequency, the conductivity minima generally occur at frequencies slightly bigger than the values of $\omega_p$ satisfying the commensurability criteria. The dependence on $\zeta$ (pump strength) saturates rapidly as $\zeta$ is increased above $1$.

From these results we conclude that a system can exhibit a negative conductivity when: \textit{(i)} the DoS is on average an increasing function of $\varepsilon$ in the region of equilibrium occupied states. \textit{(ii)} The pump is strong enough to induce a sizable population inversion of the electrons. \textit{(iii)} The phonon frequency $\omega_p$ is roughly commensurate with a relevant energy scale in the density of states, e.g. the distance from the Fermi level to the edges of the band or to a minimum of $\tilde{\sigma}$.

In Fig. 2d and 2e we report the dependence of $\sigma(\omega_p)$ on temperature and dc field, for the trial DoS of Fig. 1. We find that the negative conductivity is suppressed at high temperatures (Fig. 2d) and by an electric field (Fig. 2e). In particular, Joule heating dominates the entropy production at high fields, so $\sigma(E)$ must become positive as $E$ increases: thus if $\sigma(E=0)<0$, there exists a field $E^{\star}$ for which the conductivity vanishes, $\sigma(E^{\star})=0$. The value of $E^{\star}$ is set by the scattering length $v_F\sqrt{\tau_{\textrm{tr}}/\Gamma_{in}}$ and depends on the details of the system. Roughly, $\sigma>0$ when either $\St{in}$ or $\St{E}$ are large enough to smooth out the local population inversion, i.e. when the effective temperature gets of the order of $\omega_p$, or $E^{\star}\sim\sqrt{\omega_p\Gamma_{in}/\tau_{\textrm{tr}}}/ev_F$. This rough estimate agrees with Fig. 2e.

\textit{Short photoexcitation pulse} - We now consider a short pump pulse and study the subsequent evolution of $f$ and $\zeta$. We show that the system may develop a transient NAC state that persists after the drive is switched off. We describe the pump pulse with a characteristic strength $\zeta_0$ \footnote{$\zeta_0$ takes into account all the details about the pump fluence, polarization and coupling to the phonon mode. We use reasonable values of $\zeta_0$, which are roughly estimated by assuming that all the energy of the pump is absorbed by the phonon mode at energy $\omega_p$; this results in $\zeta_0\sim10$ for fluences $\sim1\um{mJ/cm^2}$ and pump penetration lengths $\sim100\um{nm}$.} and a duration $\tau_{\textrm{pulse}}$; we consider a pulse much shorter than the relaxation time, so that the time scales involved are well separated. We also assume that the inelastic scattering is small $\St{el}\gg\St{in}$ (or $\Gamma_{in}\ll\Gamma_{eph}$).

\begin{figure}[t]
\centering
\includegraphics[width=0.96\columnwidth]{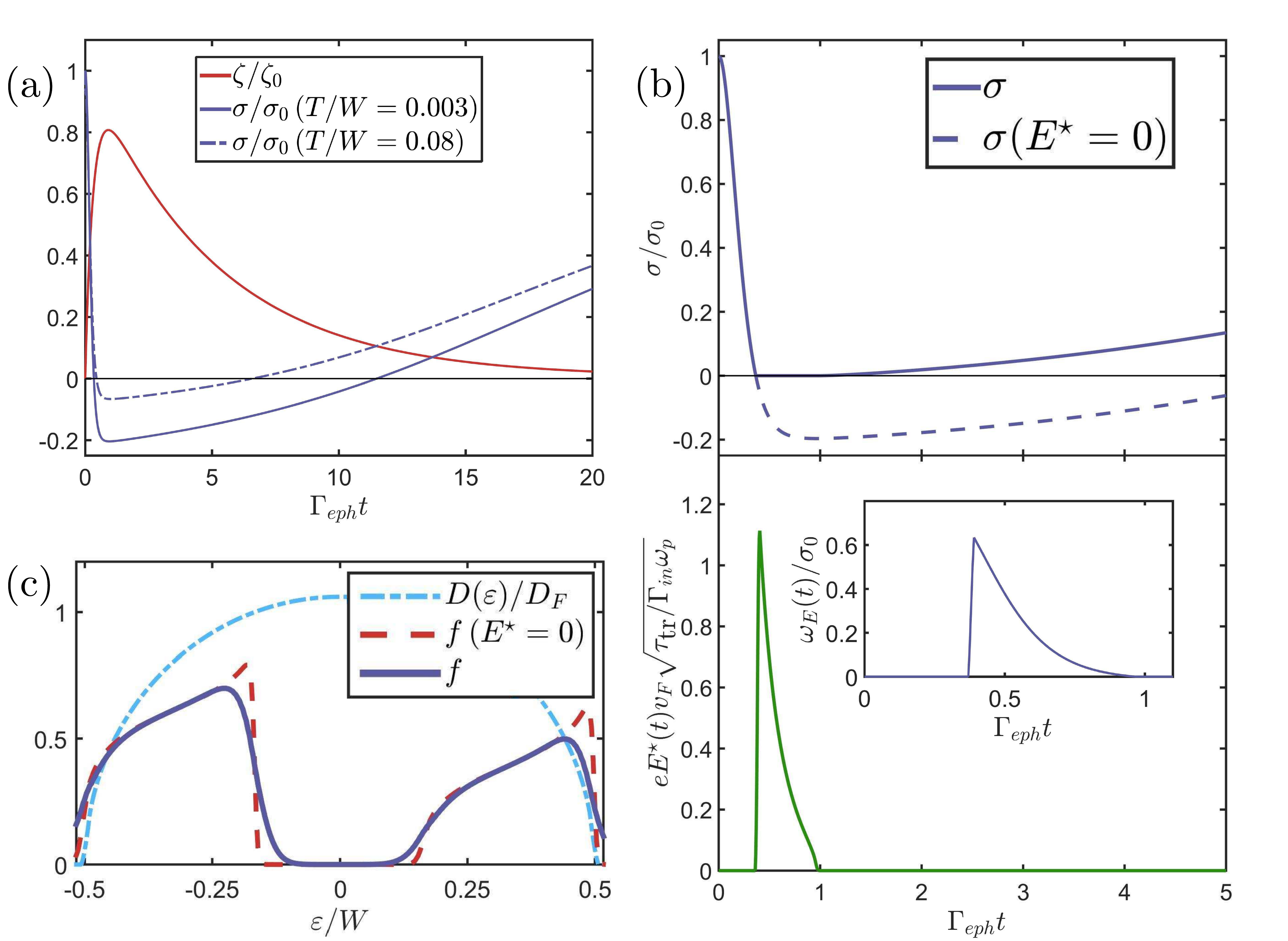}
\caption{\footnotesize{(a) Plot of $\sigma/\sigma_0$ (blue) and $\zeta/\zeta_0$ (red) as function of time $\Gamma_{eph}t$ at 1/3-filling, $\zeta_0=10$, $\omega_p/W=2/3\um{eV}$, $T/W=0.003$ (solid) and $T/W=0.08$ (dashed) for $\Gamma_{eph}\tau_{\textrm{pulse}}=0.3$ and $\Gamma_{eph}\tau_{\textrm{ph}}=5$. (b) Upper panel: plot as function of time $\Gamma_{eph}t$ of $\sigma/\sigma_0$ (blue) for the two scenarios with (solid) and without (dashed) considering the $\sigma<0$ instability; bottom panel: plot as function of time of $E^{\star}(t)$, for the same parameters as (a); the inset shows the parameter $\omega_E(t)$ normalized to $\sigma_0$. (c) Plot of $D(\varepsilon)$ and of the distribution $f$ at $\Gamma_{eph}t=5$; the red dashed curve refers to the scenario with no $\sigma<0$ instability ($E^{\star}=0$), while the solid blue curve takes the instability into account. We used the trial DoS from Fig. 1 and modeled the pulse as a decaying exponential $I_p(t)=\zeta_0\tau_{\textrm{pulse}}^{-1}e^{-t/\tau_{\textrm{pulse}}}$ for $t>0$.}}
\label{F3}
\end{figure}
We solve numerically Eqs. \eqref{KEf}-\eqref{STE} for the trial DoS of Fig. 1 and plot the behavior of $\sigma$ and $\zeta$ as function of time for $\Gamma_{eph}\tau_{\textrm{ph}}=5$ in Fig. 3a. After the pump is switched on, $\zeta(t)$ grows rapidly ($t\sim\tau_{\textrm{pulse}}$) and the system develops a negative $\sigma$ ($t\sim\Gamma_{eph}^{-1}$); $\zeta$ then relaxes back to equilibrium ($t\sim\tau_{\textrm{ph}}$) and $\sigma$ returns positive. The NAC state occurs if it can develop before the system relaxes, i.e. if $\Gamma_{eph}^{-1}\ll\tau_{\textrm{ph}}$; its lifetime is $\sim\tau_{\textrm{ph}}\gg\tau_{\textrm{pulse}}$, showing the persistence of the NAC state long after the driving pulse is removed, and exhibits a slight decrease at higher $T$ or lower $\zeta_0$, as expected.

We can see from Eq. \eqref{Kez} why the relaxation timescale is $\sim\tau_{\textrm{ph}}$ and not $\Gamma_{eph}^{-1}$. This occurs because, after an initial energy transfer from the phonon mode to the electrons, the system attains an approximate steady state in which there is no further energy transfer between electrons and phonon, because $f$ is $\omega_p$-periodic leading to $\St{el}\approx0$ and $\St{ph}\approx0$. In this situation the phonon mode can relax either through the other phonons (scale $\sim\tau_{\textrm{ph}}$) or indirectly because of the inelastic scattering of the electrons (scale $\sim\Gamma_{in}^{-1}$); in our framework, both these timescales are longer than $\Gamma_{eph}^{-1}$, leading to a rather long lived non-equilibrium state.

The results in Fig. 3a neglect the instability associated to a negative conductivity: for $\sigma<0$ any charge fluctuation grows exponentially with a characteristic time $\tau_M^{-1}=4\pi|\sigma|$ \cite{CMA,IA:AJM}; for a typical metal $\tau_M\lesssim1\um{fs}$. This time is much smaller than the typical values of $\tau_{\textrm{pulse}}$, so we can assume that the system instantaneously tunes itself to a state with a spontaneous polarization $|\vec E|=E^{\star}$ such that $\sigma(E^{\star})=0$. We take into account the instability by including the contribution of Eq. \eqref{STE} to the collision integral, with $E^{\star}(t)$ chosen so that if the solution of Eq. \eqref{KEf} predicts $\sigma(t)<0$, $\sigma(t,E^{\star}(t))=0$.

In Fig. 3b we plot $\sigma(t)$ and $E^{\star}(t)$; the value of the field is comparable with the steady state values found previously in Fig. 2e. The field grows very rapidly in a short time $\sim\tau_M$ and then decays following the relaxation of $\zeta(t)$; notice that $E^{\star}(t)$ goes to zero in a finite time and with a non zero derivative, because the conductivity turns back positive when $\zeta(t)$ decays below a certain threshold. This internal field does no affect the decay of $\zeta$, but smooths out the regions of inverted population in $f$, as observed in Fig. 3c: the non-equilibrium distributions at equal times are compared for the cases $E^{\star}=0$ and $E^{\star}\neq0$ finding a weakening of the local population inversions. This leads to a faster relaxation towards equilibrium, so that the zero conductivity state has a shorter lifetime than the NAC state (Fig. 3b).

Finally to make a connection to the phenomenological analysis of Ref. \cite{CMA}, we estimate the parameter $\omega_E$, i.e. the sensitivity of entropy production to perturbations of the total energy. Notice that in Ref. \cite{CMA} the total energy of the system is conserved after the pulse, while in this paper we allow for energy relaxation through $\St{in}$ and $\tau_{\textrm{ph}}$. Therefore, although the connection would be technically imprecise, we can still estimate $\omega_E$ as the derivative of the Joule heating contribution to entropy production with respect to fluctuations of the electric field energy: $\omega_E\sim\partial\sigma/\partial E^2|_{E^{\star}}(E^{\star})^2$; $\omega_E(t)$ depends on time and goes to zero as $E^{\star}(t)\rightarrow0$, see inset in Fig. 3b.

\textit{Conclusions} - We have studied a minimal microscopic model for the transient conductivity of a photoexcited metal, in which the pump drives a strong non-equilibrium phonon distribution, that may induce an inverted electron population.

We found the conditions for the occurrence of the population inversion and studied the dynamics of this transient state, considering the relaxation of phonons and electrons. We found that for certain pump energies (dependent on the band structure and the doping level), the photoexcited system develops an absolute negative conductivity state. Ideal systems that may exhibit such state have electron-phonon coupling strong enough so that the related scattering time is faster than the relaxation time of the system; they also have a commensurate ratio between phonon frequency and bandwidth, which is easily achieved in the case of narrow bandwidth and/or high frequency phonons. The negative conductivity state is unstable and evolves into a state with zero conductivity and a spontaneous electric polarization. We showed that this transient state persists even after the pump has been removed and that the spontaneous electric field does not immediately destroy the zero conductivity state, but rather reduces its lifetime.

\textit{Acknowledgements} - Support was provided by the Basic Energy Sciences Division of the Office of Science of the United States Department of Energy under Grant No. DE-SC0018218 (A.M. and G.C.) and by the Simons Foundation (I.A.).

\bibliographystyle{apsrev4-1}
\bibliography{bibnegsig}

\onecolumngrid

\pagebreak[4]

\begin{center}
\textbf{\Large{Supplemental material}}
\end{center}

\section{Hamiltonian}\label{ham}

We consider the Hamiltonian
\begin{equation}\label{Ham}
H=\sum_{\alpha,\vk}\varepsilon_{\vk,\alpha}
c^{\dag}_{\vk,\alpha}c_{\vk,\alpha}+\sum_{\vq}\omega_pa^{\dag}_{\vq}a_{\vq}+
\sum_{\mathbf k,\mathbf q,\alpha,\beta}M_{\mathbf q}^{\alpha\beta}(a^{\dag}_{-\vq}+a_{\vq})
c^{\dag}_{\vk,\alpha}c_{\mathbf{k-q},\beta}+H_{\textrm{anh}},
\end{equation}
where $c^{\dag}_{\vk,\alpha}$ creates an electron with momentum $\vk$ and energy $\varepsilon_{\vk,\alpha}$ on the band $\alpha$, $a^{\dag}_{\vq}$ creates a phonon with momentum $\vq$, $\omega_p$ is the energy of the phonon (assumed dispersionless), $M_{\vq}^{\alpha\beta}$ is the electron-phonon interaction. The anharmonic term $H_{\textrm{anh}}$ for the phonons is discussed in section \ref{self}.

In our specific calculations in Fig. 2c in the main text, we use $\varepsilon_{\vk,\alpha}$ for a tight binding model with hopping given by $t_{\alpha,\beta}(\vk)$, where $t_{xx}(\vk)=t_1\cos(k_x/2)\left[\cos(k_y/2)+\cos(k_z/2)\right]+t_2\cos(k_y/2)\cos(k_z/2)$, $t_{xy}(\vk)=t_3\sin(k_x/2)\sin(k_y/2)$ and cyclical permutations over the indices $x,\,y,\,z$, with $t_1=-100\um{meV}$, $t_2=30\um{meV}$ and $t_3=200\um{meV}$, chosen to qualitatively reproduce the density of states structure of K$_3$C$_{60}$.

We assume that the momentum relaxation for electron and phonons is much faster than all the other processes occurring in the system: we will average over the electron momentum later on, while the dispersionless phonon approximation means that we do not need to indicate the phonon momentum explicitly.

We define the Keldysh Green functions for electrons and phonons as:
\begin{gather}\label{Gel}
G_{\vk,\alpha}^<(t,t')=i\langle c^{\dag}_{\vk,\alpha}(t)c_{\vk,\alpha}(t')\rangle; \qquad G_{\vk,\alpha}^>(t,t')=-i\langle c_{\vk,\alpha}(t)c^{\dag}_{\vk,\alpha}(t')\rangle;\\
G_{\vk,\alpha}^R(t,t')=-i\theta(t-t')\langle\{c_{\vk,\alpha}(t),c^{\dag}_{\vk,\alpha}(t')\}\rangle;\qquad G_{\vk,\alpha}^A(t,t')=i\theta(t'-t)\langle\{c_{\vk,\alpha}(t),c^{\dag}_{\vk,\alpha}(t')\}\rangle;\\
\label{Gph}\D^<_{a^{\dag}a}(t,t')=-i\langle a^{\dag}_{\vq}(t)a_{\vq}(t')\rangle; \qquad \D^>_{a^{\dag}a}(t,t')=-i\langle a_{\vq}(t)a_{\vq}^{\dag}(t')\rangle;\\
\D^R_{a^{\dag}a}(t,t')=-i\theta(t-t')\langle[a_{\vq}(t),a_{\vq}^{\dag}(t')]\rangle;\qquad \D^A_{a^{\dag}a}(t,t')=i\theta(t'-t)\langle[a_{\vq}(t),a_{\vq}^{\dag}(t')]\rangle.
\end{gather}

We also consider $\D_{aa^{\dag}}$, obtained by exchanging the order of $a$ and $a^{\dag}$ and related to $\D_{a^{\dag}a}$ by $\D^{</>}_{aa^{\dag}}(t,t')=\D^{>/<}_{a^{\dag}a}(t',t)$ and
$\D^{R/A}_{aa^{\dag}}(t,t')=\D^{A/R}_{a^{\dag}a}(t',t)$. Note that our convention for the phonon Green function is different from the usual one, which involves an expectation of the coordinate operator $\sim(a^{\dag}+a)$). The reason to treat the $\D_{aa^{\dag}}$ and $\D_{a^{\dag}a}$ phonon Green functions separately is that the generalized Kadanoff-Baym ansatz (GKBA) employed later can only be applied separately to $\D^<_{a^{\dag}a}$ and to $\D^<_{aa^{\dag}}$.

\section{Quantum kinetic equation}\label{QKE}

We derive a quantum kinetic equation starting from the Dyson equation for the Green functions. To do so we consider interactions that relax both momentum and energy and assume that the first ones are stronger. We then assume that the non-equilibrium properties evolve more slowly than the basic electronic timescales. Finally we will apply the generalized Kadanoff-Baym ansatz to get a kinetic equation for the distribution of electrons and phonons.

\subsection{Phonon dynamics}

We start from phonons, for which the Dyson equations read
\begin{gather}\label{Keq}
(i\partial_t-\omega_p)\D^R(t,t')=\delta(t-t')+(\Sigma_{ph}\circ \D)^R(t,t');\qquad\quad(i\partial_t-\omega_p)\D^<(t,t')=(\Sigma_{ph}\circ \D)^<(t,t'),
\end{gather}
where the $\circ$ indicates the convolution in time and $\Sigma_{ph}$ is the phonon self-energy due to the interaction with the electrons and the other phonons. Here we may consider only the $a^{\dag}a$ component and so drop the subscript in $\D$.

The standard procedure is to take the hermitian conjugate of Eq. \eqref{Keq}, sum (for $\D^R$) or subtract (for $\D^<$) the two equations and pass to the coordinates $\tau=t-t'$ and $T=\frac12(t+t')$.

\begin{gather}\label{KtTr}
(i\partial_{\tau}-\omega_p)\D^R(\tau,T)=\delta(\tau)+\frac12\int d\tau'[\Sigma_{ph}^R(\tau_1,T_1)\D^R(\tau_2,T_2)+\D^R(\tau_1,T_1)\Sigma_{ph}^R(\tau_2,T_2)];\\
\label{KtTl}i\partial_T\D^<(\tau,T)=\int d\tau'[\Sigma_{ph,1}^R\D^<_2+\Sigma_{ph,1}^<\D^A_2
-\D^R_1\Sigma_{ph,2}^<-\D^<_1\Sigma_{ph,2}^A],
\end{gather}
where $\D_{1/2}\equiv\D(\tau_{1/2},T_{1/2})$, with $\tau_{1/2}=\tau/2\mp\tau'$ and $T_{1/2}=T\pm\tau_{1/2}/2$.

The dynamics associated with the relative time argument $\tau$ (scale of fs $\sim1/\omega_p$) is much faster than the dynamics associated with the ``center of mass'' time $T$ (scale of ps). Thus we assume $\tau\ll T$ and argue that the relevant $\tau'$ in the collision integral are much smaller than $T$, since the $e^{i\omega_p\tau'}$ terms in the integrand oscillate very rapidly and give a vanishing contribution when $\tau'$ becomes of order of $T$. Then $T_{1/2}\approx T$ and we Fourier transform Eq. \eqref{KtTr}-\eqref{KtTr} with respect to $\tau$, obtaining
\begin{gather}\label{DR}
(\omega-\omega_p)\D^R(\omega,T)=1+\Sigma_{ph}^R(\omega,T)\D^R(\omega,T)\,\,\,
\Rightarrow\,\,\,\D^R(\omega,T)=\frac1{\omega-\omega_p-\Sigma_{ph}^R(\omega,T)};\\
\label{Dl}i\partial_T\D^<(\omega,T)=
\D^<(\omega,T)[\Sigma_{ph}^R(\omega,T)-\Sigma_{ph}^A(\omega,T)]+
\Sigma_{ph}^<(\omega,T)[\D^A(\omega,T)-\D^R(\omega,T)].
\end{gather}

Equation \eqref{DR} is the usual expression for the retarded Green function, although there is now a time dependence in $T$ and the self-energy will be different out of equilibrium.

We now apply the generalized Kadanoff-Baym ansatz (GKBA) to Eq. \eqref{Dl}, that allows to write the non diagonal elements (in $\tau$) of the lesser Green function in terms of the diagonal part, which is proportional to the phonon distribution, here denoted by $\zeta+f_B$, where $\zeta$ is the non equilibrium part and $f_B$ the Bose distribution at energy $\omega_p$: $\D^<(\tau\gtrless0,T)=\pm i\D^{R/A}(\tau,T)\D^<(0,T-|\tau|/2)\approx\mp \D^{R/A}(\tau,T)(f_B+\zeta)$. Simplifying, we obtain
\begin{gather}
\label{GKBAw}
\D^<(\omega,T)=(\D^R(\omega)-\D^A(\omega))(f_B+\zeta)=
2i\textrm{Im}\D^R(\omega,T)(f_B+\zeta);\quad \D^>(\omega,T)=2i\textrm{Im}\D^R(\omega,T)(1+f_B+\zeta).
\end{gather}

From Eqs. \eqref{DR} we observe that for small self energies the imaginary part of the retarded Green function is very peaked around $\omega_p$ and can be approximated by a delta function: $-2i\textrm{Im}\D^R(\omega,T)\approx2\pi i\delta(\omega-\omega_p)$. We thus integrate Eq. \eqref{Dl} over $\omega$ and obtain a quantum kinetic equation for $\zeta$. Since there is no ambiguity anymore, we use the notation $t$ to indicate $T$ from now on:
\begin{equation}\label{keqz}
\partial_t\zeta=i\Sigma_{ph}^<(\omega_p,t)+i(f_B+\zeta)
[\Sigma_{ph}^A(\omega_p,t)-\Sigma_{ph}^R(\omega_p,t)].
\end{equation}

\subsection{Electron dynamics}

A similar procedure is applied to the electrons, for which the effect of a dc electric field is also considered. We assume that the magnitude of the field is small enough to not cause any modification to the band structure of the system, so that its effects on the retarded Green functions can be neglected, but strong enough to affect the distribution function. The equations are then:
\begin{gather}\label{GR}
(\varepsilon-\varepsilon_{\vk,\alpha})G^R_{\vk,\alpha}(\varepsilon,t)=
1+\Sigma^R_{\vk,\alpha}(\varepsilon,t)G^R_{\vk,\alpha}(\varepsilon,t)\,\,\,
\Rightarrow\,\,\,G^R_{\vk,\alpha}(\varepsilon,T)=\frac1{\varepsilon-\varepsilon_{\vk,\alpha}-\Sigma^R_{\vk,\alpha}(\varepsilon,t)};\\
\label{Gl}i[\partial_t+e\vec E\cdot(\vec{\nabla}_{\vk}+\vec v_{\vk,\alpha}\partial_{\varepsilon})]G^<_{\vk,\alpha}(\varepsilon,t)=\St{}\{G^<\};\\
\label{St}\St{}\{G^<\}=G^<_{\vk,\alpha}(\varepsilon,t)[\Sigma^R_{\vk,\alpha}(\varepsilon,t)-\Sigma^A_{\vk,\alpha}(\varepsilon,t)]+
\Sigma^<_{\vk,\alpha}(\varepsilon,t)[G^A_{\vk,\alpha}(\varepsilon,t)-G^R_{\vk,\alpha}(\varepsilon,t)].
\end{gather}
We have defined the collision integral $\St{}\{G^<\}$. Similarly to Eq. \eqref{DR} the retarded Green function in Eq. \eqref{GR} has an equilibrium-like structure with a modified and time-dependent self-energy $\Sigma_{\vk\,\alpha}(\varepsilon,t)$. ; equation \eqref{Gl} is the starting point to derive the expression for the conductivity $\sigma$ and the kinetic equation for the electron distribution function $f$.

The electric current is given by
\begin{equation}\label{JG}
\vec j(t)=e\int\frac{d\varepsilon}{2\pi i}\sum_{\vk,\alpha}\vec v_{\vk,\alpha}G^<_{\vk,\alpha}(\varepsilon,t)
\end{equation}
Since the velocity $\vec v_{\vk,\alpha}=\vec{\nabla}_{\vk}\epsilon_{\vk,\alpha}$ is odd in momentum, only the odd part of $G^<_{\vk,\alpha}$ contributes to the current.

The collision integral in Eq. \eqref{St} can be written as the sum of a scattering term $\St{p}\{G^<\}$ which relaxes momentum but not energy, plus an energy relaxation term $\St{en}\{G^<\}$; usually the momentum relaxation is much faster than the energy relaxation (i.e. $\St{p}\{G^<\}\gg\St{en}\{G^<\}$) so that the momentum anisotropy of $G^<$ will be small compared to the momentum averaged value of $G^{<}$. We can therefore write the GKBA ansatz as
\begin{gather}\label{GAF}
G^<_{\vk,\alpha}=2\pi i\textrm{Im}G^R_{\vk,\alpha}(\varepsilon,t)f(\varepsilon,t)+\delta G_{\vk,\alpha}(\varepsilon,t)\equiv A_{\vk,\alpha}(\varepsilon,t)f(\varepsilon,t)+\delta G_{\vk,\alpha}(\varepsilon,t);\\
\label{FF}G^<(\varepsilon,t)\equiv\frac1N\sum_{\vk,\alpha}G^<_{\vk,\alpha}(\varepsilon,t);\qquad
\sum_{\vk}\delta G_{\vk,\alpha}=0;\qquad \vec F(\varepsilon,t)\equiv\sum_{\vk,\alpha}\vec v_{\vk,\alpha}G^<_{\vk,\alpha}(\varepsilon,t)
\end{gather}
and solve Eqs. \eqref{Gl} and \eqref{St} doing perturbation theory in terms of $\vec F$. In Eqs. \eqref{GAF}-\eqref{FF}, $f$ is the distribution function, $A_{\alpha}$ is the spectral weight of the band $\alpha$ and $N$ is the number of $\vk$ states.

We act with the operator $\sum_{\vk}\vec v_{\vk,\alpha}$ on Eq. \eqref{Gl}. We neglect the energy relaxation terms surviving after the sum, as they are much smaller than the momentum relaxation, and write the momentum relaxation self energy in terms of an interaction matrix element $\Sigma^a_{\alpha}(\varepsilon,T)=\sum_{\vq}V^2_{\vq}G_{\vq,\alpha}^a(\varepsilon,T)$ within a self-consistent Born approximation:
\begin{gather*}
\sum_{\vk,\alpha}\vec v_{\vk,\alpha}\St{p}\{G^<\}=\sum_{\vk,\alpha}\left[\vec v_{\vk,\alpha} (G^A_{\vk,\alpha}(\varepsilon,t)-G^R_{\vk,\alpha}(\varepsilon,t))\sum_{\vq}V^2_{\vq}
G^<_{\vq,\alpha}(\varepsilon,t)-\vec v_{\vk,\alpha} G^<_{\vk,\alpha}(\varepsilon,t)\sum_{\vq}V^2_{\vq}(G^A_{\vq,\alpha}(\varepsilon,t)-
G^R_{\vq,\alpha}(\varepsilon,t))\right]=\\
=-\vec F(\varepsilon,t)\sum_{\vq}V^2_{\vq}(G^A_{\vq,\alpha}(\varepsilon,t)-
G^R_{\vq,\alpha}(\varepsilon,t))\approx-i\vec F(\varepsilon,t)/\tau_{\textrm{tr}}
\end{gather*}
where the $\vk$ sum in the first term of the right hand side vanishes because of the odd parity of $\vec v_{\vk}$ and we have defined the transport scattering time $\tau_{\textrm{tr}}$.

The $t$ evolution of the system happens on a timescale larger than $\tau_{\textrm{tr}}$, i.e. $\tau_{\textrm{tr}}\partial_t\vec F\ll \vec F$, and Eq. \eqref{Gl} becomes
\begin{equation}\label{Jeq1}
\vec F(\varepsilon,t)\approx
-e\tau_{\textrm{tr}}\sum_{\vk,\alpha}\vec v_{\vk,\alpha}\vec E\cdot(\vec{\nabla}_{\vk}+\vec v_{\vk,\alpha}\partial_{\varepsilon})G^<_{\vk,\alpha}(\varepsilon,t).
\end{equation}

From Eq. \eqref{Jeq1} we see that the condition $\delta G_{\vk}\ll G^<$ holds for fields satisfying the condition $eEv_F\tau_{\textrm{tr}}/\epsilon_F\ll1$, with $v_F$ the Fermi velocity and $\epsilon_F$ the Fermi energy; we can thus use Eq. \eqref{GAF} and neglect $\delta F_{\vk}$. From Eq. \eqref{GR} we derive $\nabla_{\vk}A_{\alpha}(\vk,\varepsilon)=-\vec v_{\vk,\alpha}\partial_{\varepsilon}A_{\alpha}(\vk,\varepsilon)$ and $(\vec{\nabla}_{\vk}+\vec v_{\vk,\alpha}\partial_{\varepsilon})A_{\alpha}(\vk,\varepsilon)f(\varepsilon,T)=\vec v_{\vk,\alpha}A_{\alpha}(\vk,\varepsilon)
\partial_{\varepsilon}f(\varepsilon,T)$ and find
\begin{equation}\label{F2}
\vec F(\varepsilon,t)=e\sum_{\vk,\alpha}\vec v_{\vk,\alpha}\vec E\cdot\vec
v_{\vk,\alpha}\tau_{\textrm{tr}}A_{\alpha}(\vk,\varepsilon)(-\partial_{\varepsilon}f(\varepsilon,t))
\end{equation}

We substitute into Eq. \eqref{JG} and write the conductivity $\sigma$
\begin{gather}\label{sigma}
\vec j(t)=e\sum_{\vk,\alpha}\int\frac{d\varepsilon}{2\pi i}\vec F(\varepsilon,t);\qquad
\sigma(t)=\frac{e^2\tau_{\textrm{tr}}}{3}\sum_{\alpha}\int
d\varepsilon D_{\alpha}(\varepsilon)v^2_{\alpha}(\varepsilon)\left(-\frac{\partial f(\varepsilon)}{\partial\varepsilon}\right).
\end{gather}

We have used $\sum_{\vk}A_{\alpha}(\vk,\varepsilon)=2\pi iD_{\alpha}(\varepsilon)$, $\sum_{\vk}A_{\alpha}(\vk,\varepsilon)\vec v_{\vk,\alpha}^2=2\pi iD_{\alpha}(\varepsilon)v_{\alpha}^2(\varepsilon)$.

To obtain the kinetic equation for the distribution function we now sum Eq. \eqref{Gl} over $\vk$ and $\alpha$. The term involving a total derivative in $\vk$ vanishes; furthermore we consider small fields:
\begin{equation}\label{GS}
i\partial_tG^<(\varepsilon,t)+ie\vec E\cdot\partial_{\varepsilon}\vec F=\sum_{\vk,\alpha}\St{en}\{G^<\}
\end{equation}

We assume that the self energy responsible for the energy relaxation is independent of momentum and has negligible differences between the bands, i.e. $\Sigma^a_{\alpha}\approx\Sigma^a$ so that
\begin{equation}\label{approx}
\sum_{\vk,\alpha}\St{en}\{G^<\}\approx \sum_{\vk,\alpha}G^<_{\vk,\alpha}(\varepsilon,t)
[\Sigma^R(\varepsilon,t)-\Sigma^A(\varepsilon,t)]+
\Sigma^<(\varepsilon,t)
\sum_{\vk,\alpha}[G^A_{\vk,\alpha}(\varepsilon,t)-G^R_{\vk,\alpha}(\varepsilon,t)]
\end{equation}

We combine Eqs. \eqref{GS}, \eqref{approx}, \eqref{F2} and neglect the time change of $D_{\alpha}$ compared to that of $f$, obtaining
\begin{gather}\label{keqe}
\sum_{\alpha}\left(-D_{\alpha}(\varepsilon)\partial_tf(\varepsilon)+e^2\frac{\vec E^2}3\tau_{\textrm{tr}} \partial_{\varepsilon}(v^2_{\alpha}(\varepsilon)D_{\alpha}(\varepsilon)
\partial_{\varepsilon}f(\varepsilon))-iD_{\alpha}(\varepsilon)[f(\varepsilon)
[\Sigma^R(\varepsilon,t)-\Sigma^A(\varepsilon,t)]+
\Sigma^<(\varepsilon,t)]\right)=0;\\
\label{keq}\partial_tf(\varepsilon)-e^2\frac{\vec E^2}3\tau_{\textrm{tr}}\frac1{D(\varepsilon)} \partial_{\varepsilon}\left(\sum_{\alpha}v^2_{\alpha}(\varepsilon)D_{\alpha}(\varepsilon)
\partial_{\varepsilon}f(\varepsilon)\right)=-i[f(\varepsilon)
[\Sigma^R(\varepsilon,t)-\Sigma^A(\varepsilon,t)]+
\Sigma^<(\varepsilon,t)].
\end{gather}

\section{Self-energy}\label{self}

\begin{figure}
  \centering
  \includegraphics[width=\textwidth]{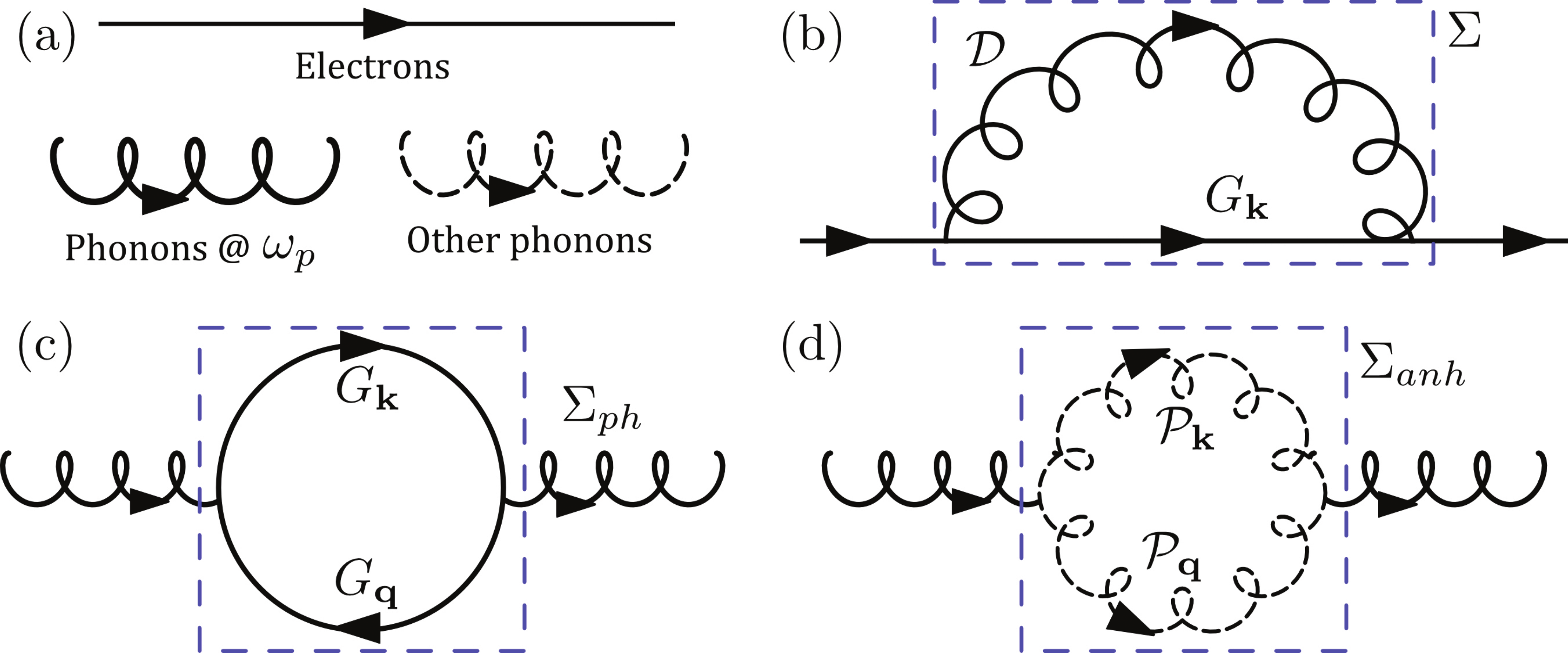}
  \caption{\scriptsize{(a) Propagators of the electrons (top), of the phonon mode at energy $\omega_p$, and of the rest of phonon branches into which the anharmonic decay occurs. (b) Self energy $\Sigma(t_1,t_2)$ diagram of the electrons due to the coupling with the phonon mode at energy $\omega_p$. (c) Self energy $\Sigma_{ph}(t_1,t_2)$ diagram of the phonon mode at energy $\omega_p$ due to the coupling with the electrons. (d) Self energy $\Sigma_{anh}(t_1,t_2)$ diagram of the phonon mode at energy $\omega_p$ due to the anharmonic coupling with other phonon branches.}}\label{Fey}
\end{figure}

Equations \eqref{keq} and \eqref{keqz} are the kinetic equations for the electrons and the phonon mode. In order to solve them it is necessary to specify the self-energy, which we calculate in a first order approximation. We assume self-consistency in the distribution functions, whose changes can be large, but use the equilibrium spectral function.

We assume the self energy of the electrons to arise from the interaction with the phonon mode and from the interaction with a thermal bath, with which the electrons exchange small amounts of energy.

We start by modeling the first, under the assumption that the interaction matrix elements are constant in momentum and equal to $M$:
\begin{gather*}
\Sigma(t_1,t_2)=\frac iN\sum_{\vk,\vq,\alpha}|M|^2G_{\vk-\vq,\alpha}(t_1,t_2)[\D_{b^{\dag}b}(t_1,t_2)+\D_{bb^{\dag}}(t_2,t_1)];\quad \Sigma^A(t_1,t_2)=\Sigma^R(t_2,t_1)^*;\\
\label{Sel}\Sigma^<(t_1,t_2)=i\sum_{\vk,\vq,\alpha}|M|^2G^<_{\vq,\alpha}(t_1,t_2)
\Big[\D^<(t_1,t_2)+\D^>(t_2,t_1)\Big];\\
\label{SeR}\Sigma^R(t_1,t_2)=\frac iN\sum_{\vk,\vq,\alpha}|M|^2\Big\{\Big[G^<_{\vq,\alpha}(t_1,t_2)+G^R_{\vq,\alpha}(t_1,t_2)\Big]
\Big[\D^R(t_1,t_2)+\D^A(t_2,t_1)\Big]+G^R_{\vq,\alpha}(t_1,t_2)\Big[\D^<(t_1,t_2)+\D^>(t_2,t_1)\Big]\Big\},
\end{gather*}

Passing to Fourier transform and using Eqs. \eqref{GKBAw}-\eqref{GAF}, we find
\begin{gather}\label{SelR}
\textrm{Re}\Sigma^R(\varepsilon)=\frac{\Gamma_{eph}}{2\pi D_0}\dashint d\varepsilon'D(\varepsilon')\left(\frac{\zeta+f_B+1-f(\varepsilon')}
{\varepsilon-\omega_p-\varepsilon'}+
\frac{\zeta+f_B+f(\varepsilon')}{\varepsilon+\omega_p-\varepsilon'}\right);\\
\label{SelI}-\textrm{Im}\Sigma^R(\varepsilon)=\frac{\Gamma_{eph}}{2D_0}
\left[D(\varepsilon-\omega_p)\left(\zeta+f_B+1-f(\varepsilon-\omega_p)\right)+
D(\varepsilon+\omega_p)\left(\zeta+f_B+f(\varepsilon+\omega_p)\right)\right];\\
\label{Selm}\Sigma^<(\varepsilon,T)=i\frac{\Gamma_{eph}}{D_0}[(\zeta+f_B) D(\varepsilon-\omega_p)f(\varepsilon-\omega_p)+
(1+\zeta+f_B)D(\varepsilon+\omega_p)f(\varepsilon+\omega_p)],
\end{gather}
where $\dashint$ is the principal value integral and $\Gamma_{eph}=2\pi|M|^2D_0$ is the associated scattering rate, with $D_0$ the average density of states that gives the right units.

The energy exchange with a thermal bath can be derived from the same equations upon performing the substitution $\omega_p\rightarrow\delta\varepsilon$ and $\zeta+f_B(\omega_p)\rightarrow f_B(\delta\varepsilon)$, with $\delta\varepsilon$ the typical exchanged energy and $f_B$ the Bose distribution. Therefore we do not write its self energy, but will directly show the formula for the scattering term.

For what concerns phonons, we have two type of interactions: the scattering off the electrons and anharmonic interactions with other phonon modes, which result in inelastic scattering. We start by treating the first one, whose self energy reads
\begin{gather*}\label{Sp}
\Sigma_{ph}(t_1,t_2)=-\frac i{N^2}\sum_{\vk,\vq,\alpha,\beta}|M|^2G_{\vk+\vq,\alpha}(t_1,t_2)G_{\vk,\beta}(t_2,t_1);\quad
\Sigma^<_{ph}(\tau,t)=-\frac i{N^2}\sum_{\vk,\vq,\alpha,\beta}|M|^2G^<_{\vk,\alpha}(\tau,t)G^>_{\vq,\beta}(-\tau,t);\\
\label{SpR}\Sigma^R_{ph}(\tau,T)=-i\frac1{N^2}\sum_{\vk,\vq,\alpha,\beta}|M|^2
[G^<_{\vk,\alpha}(\tau,t)G^A_{\vq,\beta}(-\tau,T)+G^R_{\vk,\alpha}(\tau,t)G^<_{\vq,\beta}(-\tau,t)]; \quad \Sigma_{ph}^A(\tau,t)=(\Sigma_{ph}^R)^*(-\tau,t).
\end{gather*}

Using again the GKBA expressions, we find
\begin{gather}\label{Sph}
\textrm{Re}\Sigma_{ph}^R(\omega)=\frac{\Gamma_{eph}}{2\pi}\sum_{\pm}\dashint d\varepsilon d\varepsilon'\frac{D(\varepsilon)D(\varepsilon')f(\varepsilon)
}{\varepsilon\pm\omega-\varepsilon'};\qquad-\textrm{Im}\Sigma_{ph}^R(\omega)=\frac{\Gamma_{eph}}{2}\int d\varepsilon D(\varepsilon)D(\varepsilon+\omega)\left(f(\varepsilon)-f(\varepsilon+\omega)\right);\\
\label{Sphm}\Sigma_{ph}^<(\omega,T)=-i\Gamma_{eph}\int d\varepsilon D(\varepsilon)D(\varepsilon-\omega) f(\varepsilon)(1-f(\varepsilon-\omega)).
\end{gather}

In order to treat the anharmonic relaxation, we need to specify its hamiltonian. For simplicity, we restrict ourselves to a third order expansion of the ionic potential and assume that the $\omega_p$ phonon mode has relatively high energy, so that by energy conservation only the interaction of one $\omega_p$ phonon with two other phonons has to be considered. The anharmonic Hamiltonian can then be written as
\begin{equation}\label{Hanh}
H_{\textrm{anh}}=\sum_{\vk,\vq}\sum_{r,s}M_{\textrm{anh},\vk,\vq,r,s}\Big(b_{\vk,r}b_{\vq,s}a^{\dag}_{\vk+\vq}+
b^{\dag}_{\vk,r}b^{\dag}_{\vq,s}a_{\vk+\vq}\Big),
\end{equation}
where $r,s$ indicate the phonon branches and $b$, $b^{\dag}$ are the phonon operators on these branches. The bubble that gives the anharmonic self-energy is identical to that arising from the interaction with electrons, with the difference that the legs of the phonon propagator are connected to particles, instead of particle+antiparticle. Thus we write
\begin{gather*}
\Sigma_{\textrm{anh}}(t_1,t_2)=\frac i{N^2}\sum_{\vk,\vq,r,s}|M_{\textrm{anh}}|^2\P_{\vk+\vq,r}(t_1,t_2)\P_{\vk,s}(t_1,t_2);\qquad
\Sigma_{\textrm{anh}}^<(t_1,t_2)=\frac i{N^2}\sum_{\vk,\vq,r,s}|M_{\textrm{anh}}|^2\P^<_{\vq,r}(t_1,t_2)\P^<_{\vk,s}(t_1,t_2);\\
\Sigma^R_{\textrm{anh}}(t_1,t_2)=\frac i{N^2}\sum_{\vk,\vq,r,s}|M_{\textrm{anh}}|^2\Big[\P^R_{\vq,r}(t_1,t_2)\P^<_{\vk,s}(t_1,t_2)
+\P^<_{\vq,r}(t_1,t_2)\P^R_{\vk,s}(t_1,t_2)+\P^R_{\vq,r}(t_1,t_2)\P^R_{\vk,s}(t_1,t_2)\Big],
\end{gather*}
where $\P_{\vq,r}$ is the Green function for the $b$, $b^{\dag}$ phonons at momentum $\vq$ from the branch $r$. We can define the density of states for a phonon branch $D_{\textrm{ph}}^r(\varepsilon)\equiv-\frac1{\pi}\sum_{\vq}\textrm{Im}\P_{\vq,r}^R(\varepsilon)$, with the understanding that for dispersionless phonons the density of states is just a delta function. Applying the GKBA to these phonons we obtain the lesser Green function in terms of the Bose distribution $f_B(\varepsilon)$: $\sum_{\vq}\P^<_{\vq,r}=-2\pi iD_{\textrm{ph}}^r(\varepsilon)f_B(\varepsilon)$ and calculate
\begin{gather}\label{Pm}
\Sigma^<_{\textrm{anh}}(\omega)=-i2\pi|M_{\textrm{anh}}|^2\sum_{r,s}\int d\varepsilon D_{\textrm{ph}}^r(\varepsilon)D_{\textrm{ph}}^s(\omega-\varepsilon)f_B(\varepsilon)f_B(\omega-\varepsilon);\\
\label{Pr}\textrm{Im}\Sigma^R_{\textrm{anh}}(\omega)=\pi|M_{\textrm{anh}}|^2\sum_{r,s}\int d\varepsilon
D_{\textrm{ph}}^r(\varepsilon)D_{\textrm{ph}}^s(\omega-\varepsilon)(1+ f_B(\varepsilon)+f_B(\omega-\varepsilon)).
\end{gather}

Here we have assumed that there are many phonon branches to which the $\omega_p$ mode can decay and that all these branches are well connected to thermalization mechanisms, so that their distributions are always thermal.

\section{Calculation of the scattering terms}\label{Sct}

In these section we combine the main results, namely Eqs. \eqref{keqz} and \eqref{keqe}, with the formulae for the self energies from section. In particular, from now on we only deal with objects that have one frequency and one time argument, so there would be no confusion in relabeling the average time $T$ as $t$ in order to regain consistency with the main text and be able to use the symbol $T$ for the temperature.

\subsection{Electron scattering term}\label{Esct}

We insert Eqs. \eqref{SelI}-\eqref{Selm} into the right hand side of Eq. \eqref{keqe} and find that the scattering term arising from the interaction with the phonon mode is
\begin{eqnarray}\label{Stel}
\notag\St{el}=2f(\varepsilon)\textrm{Im}\Sigma^R(\varepsilon)-i\Sigma^<(\varepsilon)&=
\frac{\Gamma_{eph}}{D_0}\Big\{&D(\varepsilon+\omega_p)\Big[(1+\zeta+f_B) f(\varepsilon+\omega_p)-(\zeta+f_B) f(\varepsilon)-f(\varepsilon+\omega_p)f(\varepsilon)\Big]+\\
\notag&&D(\varepsilon-\omega_p)\Big[(\zeta+f_B) f(\varepsilon-\omega_p)-(1+\zeta+f_B)f(\varepsilon)+
f(\varepsilon)f(\varepsilon-\omega_p)\Big]\Big\};\\
\label{KelT2}
\St{el}&=\frac{\Gamma_{eph}}{D_0}\Big\{&D(\varepsilon+\omega_p)\Big[(\zeta+f_B) (f(\varepsilon+\omega_p)-f(\varepsilon))+f(\varepsilon+\omega_p)(1-f(\varepsilon))\Big]+\\
\notag&&D(\varepsilon-\omega_p)\Big[(\zeta+f_B) (f(\varepsilon-\omega_p)-f(\varepsilon))-f(\varepsilon)(1-f(\varepsilon-\omega_p))\Big]\Big\}.
\end{eqnarray}

In order to get the inelastic scattering term we just take Eq. \eqref{Stel}, replace $\omega_p$ with $\delta\varepsilon$ and take the limit for $\delta\varepsilon\rightarrow0$. Of course $\zeta=0$ and $f_B=1/(e^{\delta\varepsilon/T}-1)\approx\frac T{\delta\varepsilon}$, with $T$ the temperature of the thermal bath
\begin{gather}\notag
\notag\St{in}=\frac{\Gamma_{in}}{\delta\varepsilon}\Big[
D(\varepsilon-\delta\varepsilon)[\frac T{\delta\varepsilon}(f_{\varepsilon-\delta\varepsilon}-f_{\varepsilon})-f_{\varepsilon}(1-f_{\varepsilon-\delta\varepsilon})]+D(\varepsilon+\delta\varepsilon)[\frac T{\delta\varepsilon}(f_{\varepsilon+\delta\varepsilon}-f_{\varepsilon)}+f_{\varepsilon+\delta\varepsilon}(1-f_{\varepsilon})]\Big]=\\
\notag=\frac{\Gamma_{in}}{\delta\varepsilon}\Big[
D(\varepsilon)[\frac T{\delta\varepsilon}(f_{\varepsilon+\delta\varepsilon}+f_{\varepsilon-\delta\varepsilon}-2f_{\varepsilon})+
f_{\varepsilon+\delta\varepsilon}-f_{\varepsilon+\delta\varepsilon}f_{\varepsilon}-f_{\varepsilon}+f_{\varepsilon}f_{\varepsilon-\delta\varepsilon})]+\\
\notag+
\delta\varepsilon\partial_{\varepsilon}D[\frac T{\delta\varepsilon}(f_{\varepsilon+\delta\varepsilon}-f_{\varepsilon-\delta\varepsilon})+
f_{\varepsilon+\delta\varepsilon}-f_{\varepsilon+\delta\varepsilon}f_{\varepsilon}+f_{\varepsilon}-f_{\varepsilon}f_{\varepsilon-\delta\varepsilon}]\Big]
\end{gather}
\vspace*{-0.5cm}
\begin{equation}\label{Stinav}
\St{in}=\frac{\Gamma_{in}}{D(\varepsilon)}\partial_{\varepsilon}\Big[
D^2(\varepsilon)[T\partial_{\varepsilon}f+f(1-f)]\Big].
\end{equation}

\subsection{Phonon scattering term}\label{Psct}

We insert Eqs. \eqref{Sph}-\eqref{Sphm} into the right hand side of Eq. \eqref{keqz} and find that the scattering term arising from the interaction with the electrons is
\begin{equation}\label{Stph}
\St{ph}=i\Sigma^<_{ph}(\omega_p)-2(\zeta+f_B)\textrm{Im}\Sigma^R_{ph}(\omega_p)=
\Gamma_{eph}\int d\varepsilon D(\varepsilon)D(\varepsilon+\omega_p)\Big[f(\varepsilon+\omega_p)(1-f(\varepsilon))+
(\zeta+f_B)(f(\varepsilon+\omega_p)-f(\varepsilon))\Big].
\end{equation}

To get the anharmonic term we repeat the same procedure with Eqs. \eqref{Pm}-\eqref{Pr} and obtain
\begin{equation*}
\St{anh}=2\pi|M_{\textrm{anh}}|^2\sum_{r,s}\int d\varepsilon D_r(\varepsilon)D_s(\omega_p-\varepsilon)\Big[f_B(\omega_p-\varepsilon)f_B(\varepsilon)-
(\zeta+f_B(\omega_p))(1+f_B(\omega_p-\varepsilon)+f_B(\varepsilon))\Big].
\end{equation*}

We observe that $f_B(\omega_p-\varepsilon)f_B(\varepsilon)-
f_B(\omega_p)(1+f_B(\omega_p-\varepsilon)+f_B(\varepsilon))=0$, because they are all equilibrium distributions, so that we can just define a phonon relaxation time $\tau_{\textrm{ph}}$ and write
\begin{equation}\label{Stanh}
\St{anh}=-\zeta2\pi|M_{\textrm{anh}}|^2\sum_{r,s}\int d\varepsilon D_r(\varepsilon)D_s(\omega_p-\varepsilon)\Big(1+f_B(\omega_p-\varepsilon)+f_B(\varepsilon)\Big)
\equiv-\frac{\zeta}{\tau_{ph}}.
\end{equation}

\section{Kinetic equations}

We finally put together the previous results in order to derive the formulae for the kinetic equations written in the main text: we insert Eqs. \eqref{Stel}-\eqref{Stinav} into Eq. \eqref{keqe} and Eqs. \eqref{Stph}-\eqref{Stanh} into Eq. \eqref{keqz}. We operate the final substitution $b\equiv f_B(\omega_p)$ in order to be consistent with the main text; we then introduce phenomenologically the effect of the pump, that is modeled as a source of phonons $I_p(t)$ in Eq. \eqref{keqz}. We obtain
\begin{equation}\label{Kin}
\partial_tf(\varepsilon)+e^2E^2\frac{\tau_{\textrm{tr}}}3\frac1{D(\varepsilon)}
\partial_{\varepsilon}\left[\sum_{\alpha}D_{\alpha}(\varepsilon)v_{\alpha}^2(\varepsilon)
\partial_{\varepsilon}f(\varepsilon)\right]=\St{el}+\St{in};\qquad\qquad \partial_t\zeta=\St{ph}+I_p(t)-\frac{\zeta}{\tau_{ph}}.
\end{equation}

\section{Effect of dispersive phonons}

In this section we consider the effect of a dispersive phonon energy on the electron collision integral. The immediate effect of a dispersive phonon is that the GKBA expression Eq. \eqref{GKBAw} for the lesser phonon Green function does not present a delta function anymore, but rather a phonon density of states $D_{\textrm{ph}}(\omega)$:
\begin{equation}\label{disp1}
\delta(\omega-\omega_p)\rightarrow D_{\textrm{ph}}(\omega)\,\,\,\Rightarrow\,\,\,-2i\textrm{Im}\D^R(\omega,T)=2\pi iD_{\textrm{ph}}(\omega)
\end{equation}

This modifies the expressions for the electron self-energy in Eq. \eqref{SelR}-\eqref{Selm}, since we find
\begin{gather}
\notag\Sigma^<(\varepsilon)=i|M|^2\int\frac{d\omega}{2\pi}\sum_{\vq,\alpha}G^<_{\vq,\alpha}(\epsilon-\omega)[\D^<(\omega)+\D^>(-\omega)]=
-|M|^2\int d\omega D(\varepsilon-\omega)[\D^<(\omega)+\D^>(-\omega)]\\
\label{SelmC}\Sigma^<(\varepsilon)=i\Gamma_{eph}\int d\omega D_{\textrm{ph}}(\omega)[(\zeta+f_B) D(\varepsilon-\omega)f(\varepsilon-\omega)+
(1+\zeta+f_B)D(\varepsilon+\omega)f(\varepsilon+\omega)];\\
\label{SelIC}-\textrm{Im}\Sigma^R(\varepsilon)=\frac{\Gamma_{eph}}{2}\int d\omega D_{\textrm{ph}}(\omega)
\left[D(\varepsilon-\omega)\left(\zeta+f_B+1-f(\varepsilon-\omega)\right)+
D(\varepsilon+\omega)\left(\zeta+f_B+f(\varepsilon+\omega)\right)\right]
\end{gather}

The self-energy reduces to Eq. \eqref{SelR}-\eqref{Selm} when the phonon density of states is a delta function (or a very peaked function) at $\omega=\omega_p$. The scattering term is modified to
\begin{eqnarray}\label{StelC}
\St{el}&=\Gamma_{eph}\int d\omega D_{\textrm{ph}}(\omega)\Big\{&D(\varepsilon+\omega)\Big[(\zeta+f_B) (f(\varepsilon+\omega)-f(\varepsilon))+f(\varepsilon+\omega)(1-f(\varepsilon))\Big]+\\
\notag&&D(\varepsilon-\omega)\Big[(\zeta+f_B) (f(\varepsilon-\omega)-f(\varepsilon))-f(\varepsilon)(1-f(\varepsilon-\omega))\Big]\Big\}.
\end{eqnarray}

The integral over different frequencies of the scattering leads naturally to a correction term with a diffusive nature; in fact, if we assume that $D_{\textrm{ph}}(\omega)$ is peaked around $\omega_p$, we can expand to the linear order in $\omega-\omega_p$, finding that $\St{el}=\St{el}^{(0)}+\delta\St{el}^{(1)}$, with $\St{el}^{(0)}$ being the collision integral given by Eq. \eqref{Stel}
\begin{eqnarray}\label{Stel1}
\delta\St{el}^{(1)}&=\Gamma_{eph}\langle\delta\omega\rangle\Big\{&\partial_{\varepsilon}D(\varepsilon+\omega_p)\Big[(\zeta+f_B) (f(\varepsilon+\omega_p)-f(\varepsilon))+f(\varepsilon+\omega_p)(1-f(\varepsilon))\Big]-\\
\notag&&\partial_{\varepsilon}D(\varepsilon-\omega_p)\Big[(\zeta+f_B) (f(\varepsilon-\omega_p)-f(\varepsilon))-f(\varepsilon)(1-f(\varepsilon-\omega_p))\Big]+\\
\notag&&D(\varepsilon+\omega_p)\partial_{\varepsilon}f(\varepsilon+\omega_p)\Big[\zeta+f_B+1-f(\varepsilon)\Big]-
D(\varepsilon-\omega_p)\partial_{\varepsilon}f(\varepsilon-\omega_p)\Big[\zeta+f_B+f(\varepsilon)\Big]\Big\}
\end{eqnarray}
where $\langle\delta\omega\rangle\equiv\int d\omega D_{\textrm{ph}}(\omega)(\omega-\omega_p)$.

We thus observe that $\delta\St{el}^{(1)}/\St{el}^{(0)}\sim\textrm{max}(\langle\delta\omega\rangle\partial_{\varepsilon}D/D,
\langle\delta\omega\rangle\partial_{\varepsilon}f)$, so that the order of the correction due to a dispersive phonon band is proportional to the ratio between the typical band width $\delta\omega$ and the typical energy scale of the electron density of states or distribution. The energy diffusion behavior in the correction arises from the terms proportional to energy derivatives of $f$ and leads to a renormalization of the temperature $T\rightarrow T_{\textrm{eff}}=T+\zeta\langle\delta\omega\rangle$.

\section{One dimensional band}

In this section we study more in detail the case of a one dimensional system with dispersion $\epsilon_k=-\frac W2\cos k$, which leads to singularities in the density of states at $\varepsilon=\pm W/2$. In fact, we obtain
\begin{equation}\label{1Dev}
D(\varepsilon)=\frac2{\pi W}\frac1{\sqrt{1-(2\varepsilon/W)^2}};\qquad\qquad v^2(\varepsilon)=\frac{W^2}4-\varepsilon^2
\end{equation}

\begin{figure}[!t]
  \centering
  \includegraphics[width=0.9\textwidth]{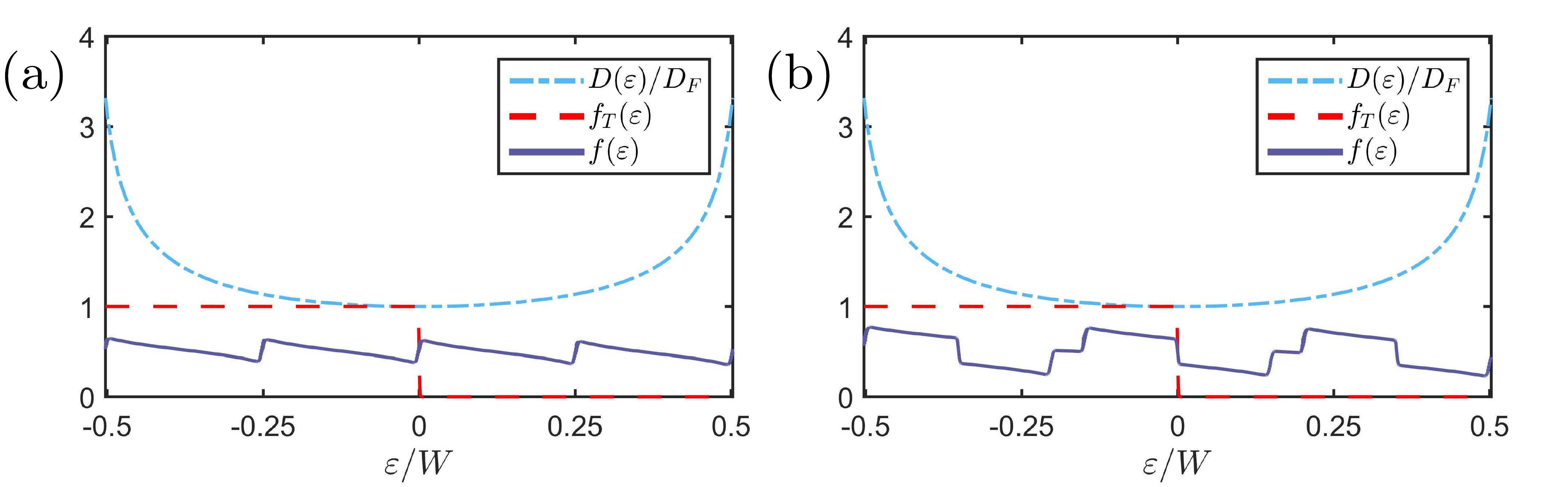}
  \caption{\scriptsize{Non-equilibrium steady state electron distribution $f$ (blue solid lines), obtained from solution of Eq. \eqref{Kin} for a steady state phonon population $\zeta=20$, initial distribution given by a Fermi-Dirac $f_T$ (red dashed) at chemical potential $\mu=0$ and temperature $T/W=0.003$, and trial DoS $D(\varepsilon)$ (cyan dashed-dotted lines) regularized at the divergences in order to avoid numerical issues. The phonon frequency is $\omega_p/W=0.25$ (a) and $\omega_p/W=0.35$ (b).}}\label{S1}
\end{figure}

We calculate the non-equilibrium distribution in the same limit used to obtain Fig. 1 in the main text: we neglect the phonon dynamics and the inelastic scattering ($\St{in}\rightarrow0$), we consider $E=0$ and $T\rightarrow0$, quench at $t=0$ the value $\zeta\gg1$ and evolve the system for a time $t=10\Gamma_{eph}^{-1}$. As expected, the distribution $f$ shows upward steps at energies that match the singularities in the DoS, see Fig. 1. In particular, the upward steps are at energies $\varepsilon+j\omega_p=\pm W/2$, while the downward steps of thermal origin are at $\varepsilon+j\omega_p=\mu$; the regions in between have a gentle decreasing behavior, in contrast with the case shown in the main text. For the particular commensurate case $\omega_p/W=0.25$ (Fig. 1a), for which $\omega_p$ is commensurate with the distance from chemical potential to both band edges, the upward steps due to the DoS singularity superimpose with the downward steps of thermal origin, resulting in only small upward steps.

As an example we calculate the conductivity of such system as function of $\omega_p$ and for different values of $\zeta$. We consider two different origins of the transport scattering processes: in the first case $\tau_{\textrm{tr}}$ is assumed constant, while in the second case we consider an impurity-like scattering that results in $\tau_{\textrm{tr}}(\varepsilon)\sim1/D(\varepsilon)$. For these two limits, we find the energy dependence of $\tilde{\sigma(\varepsilon)}$:
\begin{equation}\label{s1D}
\tilde{\sigma}(\varepsilon)=\sigma_0\sqrt{1-\left(\frac{2\varepsilon}W\right)^2};\qquad
\tilde{\sigma}(\varepsilon)=\sigma_0\left(1-\frac{2\varepsilon}W\right)^2
\end{equation}

\begin{figure}[t]
  \centering
  \includegraphics[width=0.9\textwidth]{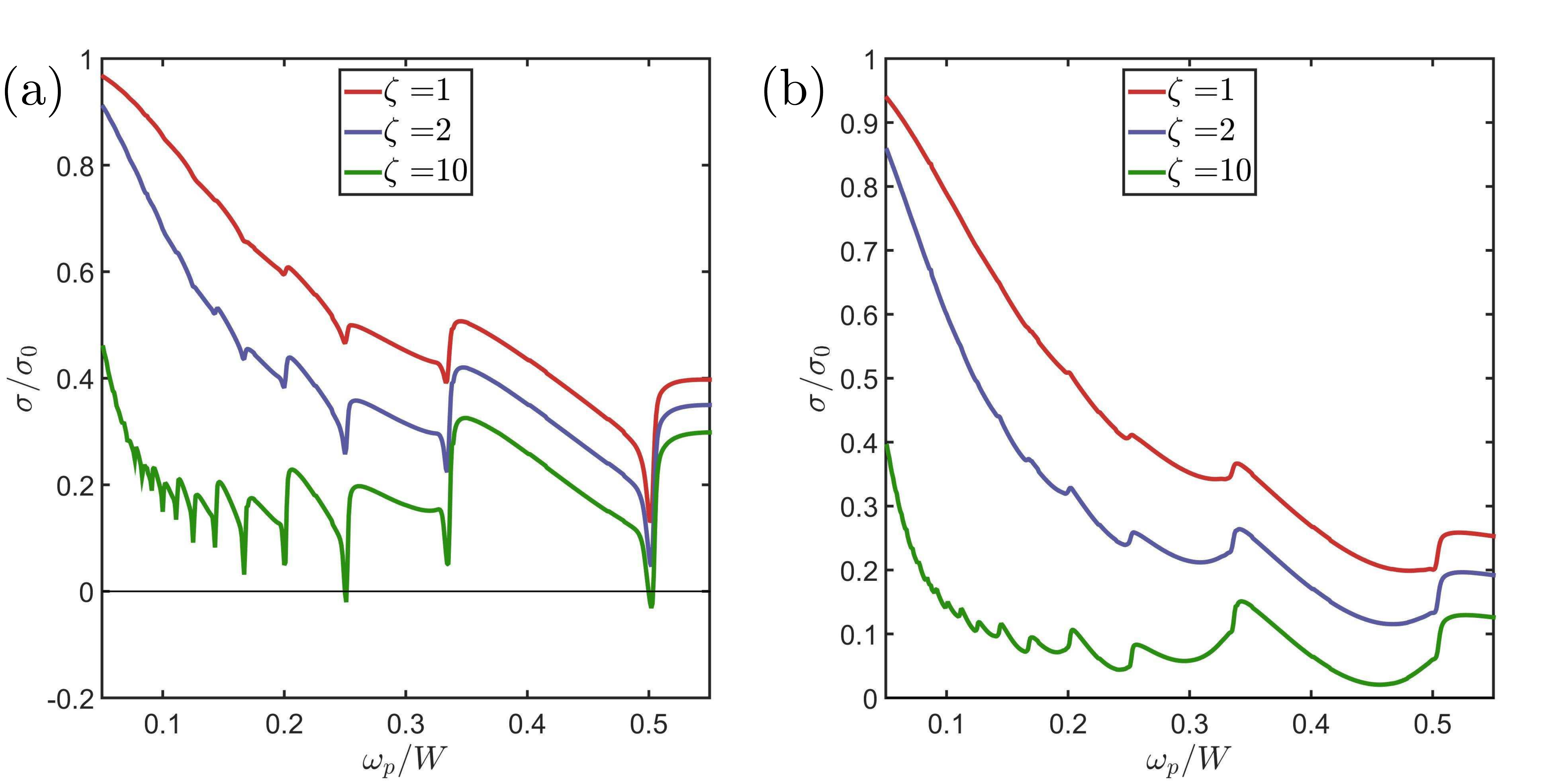}
  \caption{\scriptsize{Plot of $\sigma/\sigma_0$ as function of the normalized phonon frequency $\omega_p/W$ for three different values of $\zeta$ at $E=0$, $\St{in}=0$ and $T/W=0.003$. Panel (a) shows the case of constant $\tau_{\textrm{tr}}$ and panel (b) shows the case of impurity-like scattering with $\tau_{\textrm{tr}}\sim1/D(\varepsilon)$.}}\label{S2}
\end{figure}

As can be seen in Fig. 2, for the less realistic case of constant $\tau_{\textrm{tr}}$ the conductivity is almost always positive with only some negative values in correspondence of particularly commensurate values of $\omega_p$ such as $W/2$ and $W/4$ (remember that $\mu=W/2$). On the other hand, for the case of impurity-like scattering, the conductivity is always positive.

This brief analysis shows that a density of states that is generally increasing in the energy range from the lower band edge to the chemical potential is favorable for yielding a population inversions that leads to a negative conductivity, whereas a decreasing density of states is less favorable.

\end{document}